\documentclass[11pt]{article}

\usepackage{epsfig}
\usepackage{amsmath}
\usepackage{amscd}
\newcommand{\particle}[1] {\mathrm{#1}}

\renewcommand{\em}   {\particle{e^-}}
\newcommand{\ep}     {\particle{e^+}}
\newcommand{\el}     {\particle{e}}
\newcommand{\epem}     {\particle{e^+e^-}}
\newcommand{\nue}    {\nu_{\el}}
\newcommand{\nueb}   {\bar{\nu_{\el}}}

\newcommand{\taum}   {\tau^{-}}
\newcommand{\taup}   {\tau^{+}}

\newcommand{\lpt}    {\ell}

\newcommand{\bq}     {\particle{b}}
\newcommand{\bqb}    {\particle{\bar{\bq}}}
\newcommand{\tq}     {\particle{t}}
\newcommand{\tqb}    {\particle{\bar{\tq}}}
\newcommand{\q}      {\particle{q}}
\newcommand{\qb}     {\particle{\bar{\q}}}

\newcommand{\WB}     {\particle{W}}

\newcommand{\ZB}     {\particle{Z}}

\newcommand{\HB}     {\particle{H}}

\newcommand{\unit}[1] {\mathrm{#1}}

\newcommand{\GeV}{\unit{GeV}}

\newcommand{\fb} {\unit{fb}}

\newcommand{\ifb} {\fb^{-1}}

\newcommand{\journal}[4] {{\rm #1}~{\bf #2}\,(#3),~#4}

\newcommand{\CPC}[3]  {\journal{Comp.~Phys.~Comm.}{#1}{#2}{#3}}

\newcommand{\lsim}{\raisebox{-0.07cm}{$\,\stackrel{<}{{\scriptstyle\sim}}\,$}} 
\newcommand{\gsim}{\raisebox{-0.07cm}{$\,\stackrel{>}{{\scriptstyle\sim}}\,$}} 
\newcommand{\br}[1]{\ensuremath{\mathrm{BR}_{#1}}}
\newcommand{\rf}[1] {\ensuremath{RF_{#1}}}
\newcommand{\num}[1] {\ensuremath{N_{#1}}}
\newcommand{\prb}[1] {\ensuremath{P_{#1}}}
\newcommand{\xsec}[1]{\sigma_{#1}}
\newcommand{\mass}[1]{m_{#1}}

\begin{document}

\newcommand{\gammatot}{\Gamma_{\mathrm{tot}}}
\newcommand{\lumi}{{\cal \int\!\! L}}
\newcommand{\hww}{\HB\to\WB\WB}
\newcommand{\hzz}{\HB\to\ZB\ZB}
\newcommand{\rfww}{\rf{\hww}}
\newcommand{\rfzz}{\rf{\hzz}}
\newcommand{\brww}{\br{\hww}}
\newcommand{\brzz}{\br{\hzz}}
\newcommand{\nww}{\num{\hww}}
\newcommand{\nzz}{\num{\hzz}}
\newcommand{\pww}{\prb{\hww}}
\newcommand{\pzz}{\prb{\hzz}}
\newcommand{\eww}{\epsilon_{\hww}}
\newcommand{\ezz}{\epsilon_{\hzz}}

\rightline{LC-PHSM-2003-066}
\vspace{1cm}

\begin{center}
\boldmath
{\Huge\bf Measuring Resonance Parameters of Heavy Higgs Bosons at TESLA}
\unboldmath
\vspace{1cm}

\begin{tabular}{p{1.25cm}p{5cm}}
  \epsfig{file=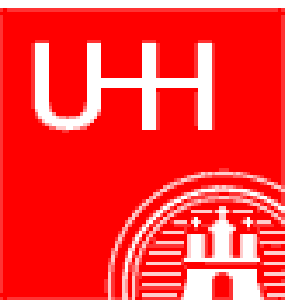,width=1.25cm} &
  \begin{tabular}[b]{l}
    Niels Meyer \\ Institute of Experimental Physics \\ 
    University of Hamburg \\ 
  \end{tabular} \\
\end{tabular}
\vspace{.5cm} 
\end{center}

\begin{abstract}
This study investigates the
potential of the TESLA Linear Collider for measuring resonance parameters of 
Higgs bosons beyond the mass range studied so far.

The analysis is based on the reconstruction of events from the Higgsstrahlung
process 
${\ep\em\to\HB\ZB}$. It is shown that the total width $\Gamma_{\HB}$, the
mass $\mass{\HB}$ and the event rate can be measured from the mass spectrum in
a model independent fit. Also, the branching ratios $\br{\hww}$ and
$\br{\hzz}$ can be measured, assuming these are the only relevant Higgs decay
modes. 

The simulation includes realistic detector effects and all relevant Standard
Model background processes. Results are given for 
$\mass{\HB}=200-320\,\GeV$ assuming $\lumi=500\,\ifb$ integrated
luminosity at collision energies of $\sqrt{s}=500\,\GeV$.  
\end{abstract}

\section{\label{sec:intro}Introduction}

During the past years many simulations have been performed to investigate the
prospects of measuring Higgs boson properties at future Linear Colliders
\cite{tdr,Abe:2001wn,Abe:2001gc}. The main 
focus was set to Higgs masses 
$\mass{\HB}$ below the $\WB\WB$-threshold which is the mass region
prefered by recent electroweak data. 

\begin{figure}[htbp]
  \center{\epsfig{file=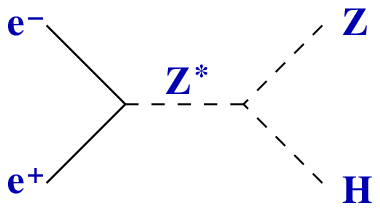,height=2.5cm}\qquad\qquad\epsfig{file=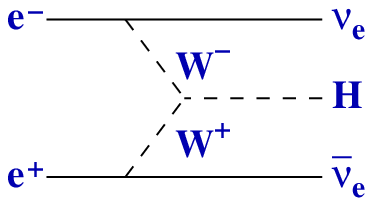,height=2.5cm}}
  \caption{\label{fig:feyn-higgs}\it
    Higgs production diagrams: Higgsstrahlung (left) and $\WB\WB$-fusion
    (right).} 
\end{figure}

\begin{figure}[htbp]
  \center{\epsfig{file=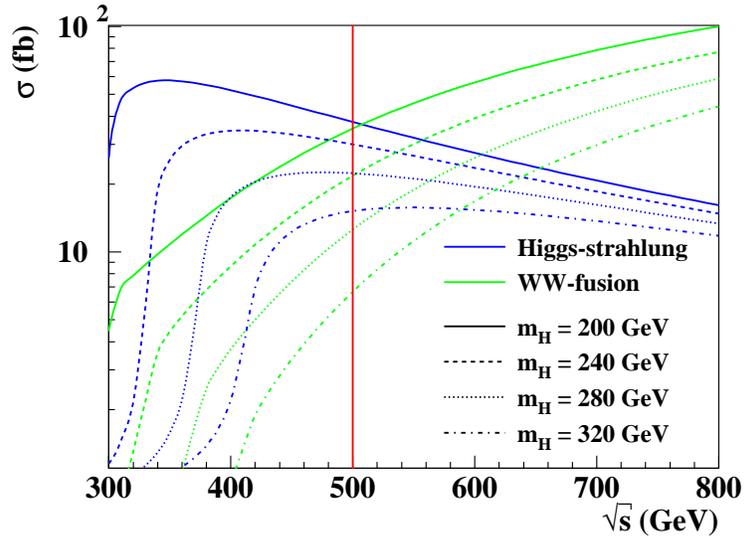,width=10cm}
  \caption{\label{fig:xsec_theory}
    \it Higgs production cross sections as predicted by the Standard
    Model. Values are calculated using {\tt HZHA} \cite{hzha}.}}
\end{figure}

In high energy $\epem$ collisions Higgs bosons can be produced in two
dominant production processes: Higgsstrahlung $\epem\to\HB\ZB$ and
$\WB\WB$-fusion $\epem\to\HB\nue\nueb$, see Fig.~\ref{fig:feyn-higgs}. For
lower collision energies $\sqrt{s}$, Higgsstrahlung with $\xsec{\HB\ZB}\sim
1/s$ dominates, while $\WB\WB$-fusion becomes the major production mode due its
$\xsec{\HB\nue\nueb}\sim\log s$ cross-section rise if $\sqrt{s}$ is large
compared to $\mass{\HB}$. The cross sections of both
processes for the considered range of Higgs masses are plotted in
Fig.~\ref{fig:xsec_theory}.  

\begin{figure}[htbp]
  \center{\epsfig{file=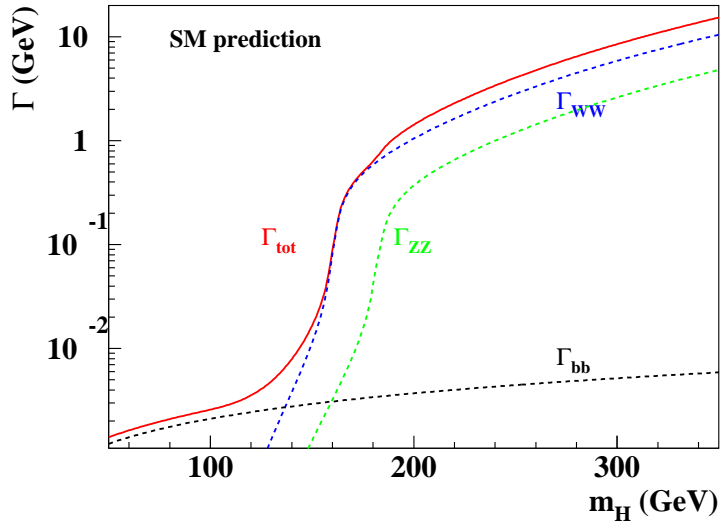,width=10cm}
  \caption{\label{fig:width_sm}
    \it Standard Model prediction for the total Higgs width
    $\gammatot$. Also shown are
    the partial widths $\Gamma_{\bq\bqb}$, $\Gamma_{\WB\WB}$ and
    $\Gamma_{\ZB\ZB}$ of the dominant decay modes. Values are calculated using
    {\tt HDECAY} \cite{hdecay}.}}
\end{figure}

Higgs bosons couple to mass and therefore decay in general to the
heaviest particles possible. In the Standard Model (SM) and
most of 
its extensions, the total Higgs decay width is expected to be very small for
Higgs masses below the $\WB\WB$-threshold. Fig.~\ref{fig:width_sm} shows the
SM prediction. Only if the Higgs
width is as large as few $\GeV$ can it be determined from the observed
Higgs lineshape. This is not possible for smaller widths due to limited
detector resolution, but the width can only be determined indirectly via the
Higgs couplings \cite{ssr_gam,nm_fus}. 

From the lineshape not only the width, but also the mass and event rate can be
determined in a model-independent way. The
signal and background processes studied are specified in
Sec.~\ref{sec:mc}. Event selection is described in Sec.~\ref{sec:sel} followed
by the methods of estimating detector resolution in Sec.~\ref{sec:det}. For
this, a separation of $\hww$ and $\hzz$ decays is necessary, which can be
interpreted as branching fraction measurements assuming that there are no
other major decay modes. The note continues with details on
the reconstruction of Higgs resonance parameters in
Sec.~\ref{sec:mhd}. Results are summarized and discussed in Sec.~\ref{sec:res}.

\section{\label{sec:mc}Signal and Background}

Higgs bosons with Standard Model couplings are studied. The parameter space of
interest ranges from $\mass{\HB}=200\,\GeV$ to $320\,\GeV$, where both cross
section and width are large enough for precision measurements. 

SM Higgs bosons in the mass range under study, decay almost
exclusively to pairs of massive gauge bosons. The final state of interest
therefore is $\epem\to\HB\,(\ZB)\to\WB\WB/\ZB\ZB\,(\ZB)$. All successive decay
modes are listed in Tab.~\ref{tab:finstat}.
\begin{table}[htbp]
\center{\begin{tabular}{|c|c||c|c||c|r|}
\hline
& & \multicolumn{2}{c||}{$\HB\to$} & SM branching & \multicolumn{1}{c|}{Events} \\
& Final state & $\WB\WB$ & $\,\;\ZB\ZB\;\,$ & fraction & \multicolumn{1}{c|}{per $500\,\ifb$} \\
\hline\hline
1. & $2\nu + X$ & $\times$ & $\times$ & 35.08\,\% & 6145$\quad$ \\
2. & $\q\q\,\q\q\,\q\q$ & $\times$ & $\times$ & 34.3\,\% & 6000$\quad$ \\
3. & $\q\q\,\q\q\,\lpt\nu$ & $\times$ & & 19.18\,\% & 3355$\quad$ \\
4. & $\lpt\lpt\,\q\q\,\q\q$ & $\times$ & $\times$ & 7.84\,\% & 1370$\quad$ \\
5. & $\lpt\lpt\,\q\q\,\lpt\nu$ & $\times$ & & 2.94\,\% & 515$\quad$ \\
6. & $\lpt\lpt\,\lpt\lpt\,\q\q$ & & $\times$ & 0.63\,\% & 110$\quad$ \\
7. & $\lpt\lpt\,\lpt\lpt\,\lpt\lpt$ & & $\times$ & 0.03\,\%& 5$\quad$ \\
\hline
\end{tabular}
\caption{\label{tab:finstat}\it
  Signal final states and their occurrence. Numbers given are estimates using
  $\br{\hww}=0.7$, $\br{\hzz}=0.3$,
  $\br{\WB\to\lpt\nu}=\br{\ZB\to\lpt\lpt}=0.3$,
  $\br{\WB\to\q\q}=\br{\ZB\to\q\q}=0.7$ and $\xsec{\HB\ZB}=35\,\fb$ as
  predicted by the SM for $\mass{\HB}=240\,\GeV$. Here, $\lpt$ is any charged
  lepton flavor and $\q$ any quark flavor kinematically allowed.}} 
\end{table}

Approximately one third of the $\HB\ZB$-events contain more than one neutrino
and thus a precise mass reconstruction is difficult. For the rest, the fully
hadronic 
final state is by far dominant. But also, this is the channel where the
largest background contributions are expected (e.g. $\epem\to\tq\tqb$). The
same is true for the next-to-dominant decay mode (hadronic $\ZB$-decay plus
semi-leptonic decay of the $\WB$-pair). On the contrary, the gold-plated
channel with six leptons in the final state is too rare for precision
measurements. Channel 4 with one leptonic $\ZB$-decay and
hadronic $\WB$- or $\ZB$-pair decays is a good compromise between signal
rate and background contamination. 

Since $\tau$-decays always include neutrinos, only
$\lpt=\el,\,\mu$ are considered to achieve the best mass reconstruction. So, from
here on lepton only means electron or muon.

All background processes with two charged leptons plus jets are
considered. They can be classified as follows:
\begin{itemize}
  \item[\bf 6f\quad] Six fermion proceses, $\ep\em\to 2\lpt \;
    4\q$, yield events with the same final state as signal events. As will
    be shown later, this is the dominant class of background processes.
  \item[\bf 4f\quad] Four fermion processes including $\ZB\ZB$-pair 
    production, $\ep\em\to 2\lpt \;2\q$. First of all, this class of
    processes is problematic because of huge cross sections. However, event
    topology differs from signal events. 
  \item[\boldmath$\tq\tqb$\unboldmath\quad] Top quark pair production, 
    $\epem\to \tq\tqb\to\bq\bqb\WB^+\WB^-$. Here, high energetic leptons might
    not only occur in $\WB$- but also in $\bq$-decays. Therefore, all decays
    $\WB\to\q\qb$ and $\WB\to\lpt\nu$ are considered. Again it is not
    event topology but large cross section which makes this backgroud possibly
    dangerous.
\end{itemize}

Other processes (e.g. $\epem\to\WB\WB\to\q\q\lpt\nu$) are expected to
be negligible due to missing isolated leptons, large missing energy or 
low mass of the hadronic system.

Signal as well as background events are generated using {\tt WHiZard
  1.22} \cite{whiz}, except $\tq\tqb$-events for which {\tt PYTHIA 6.2}
\cite{pythia} is used. Both initial state radiation ISR and beamstrahlung
\cite{circe} are taken into account. For this analysis, no significant signal
over background enhancement is expected for polarized beams, so the
possibility of beam polarization is not studied. Cuts on fermion-pair
invariant masses $\mass{\lpt\lpt}/\mass{\q\q}>10\,\GeV$ for any
lepton-/quark-pair are applied on MC level to save CPU power. These events
anyhow are far from the parameter space of interest.

Detector response is simulated using the fast Monte Carlo {\tt SIMDET 4}
\cite{simdet} which parameterizes detector performance as described in the
TESLA TDR \cite{tdr}.

\section{\label{sec:sel}Event Selection}

In all events, energy flow objects, which are classified as lepton (electron
or muon) 
are searched for. The most energetic of these objects is combined with any
other identified lepton. The pair with invariant mass closest to $\mass{\ZB}$
is selected as $\ZB\to\lpt\lpt$ candidate and removed from the event. All
other energy flow objects are forced to four jets by the Durham recombination
scheme \cite{durham}. 

\begin{figure}[htbp]
  \center{
    \epsfig{file=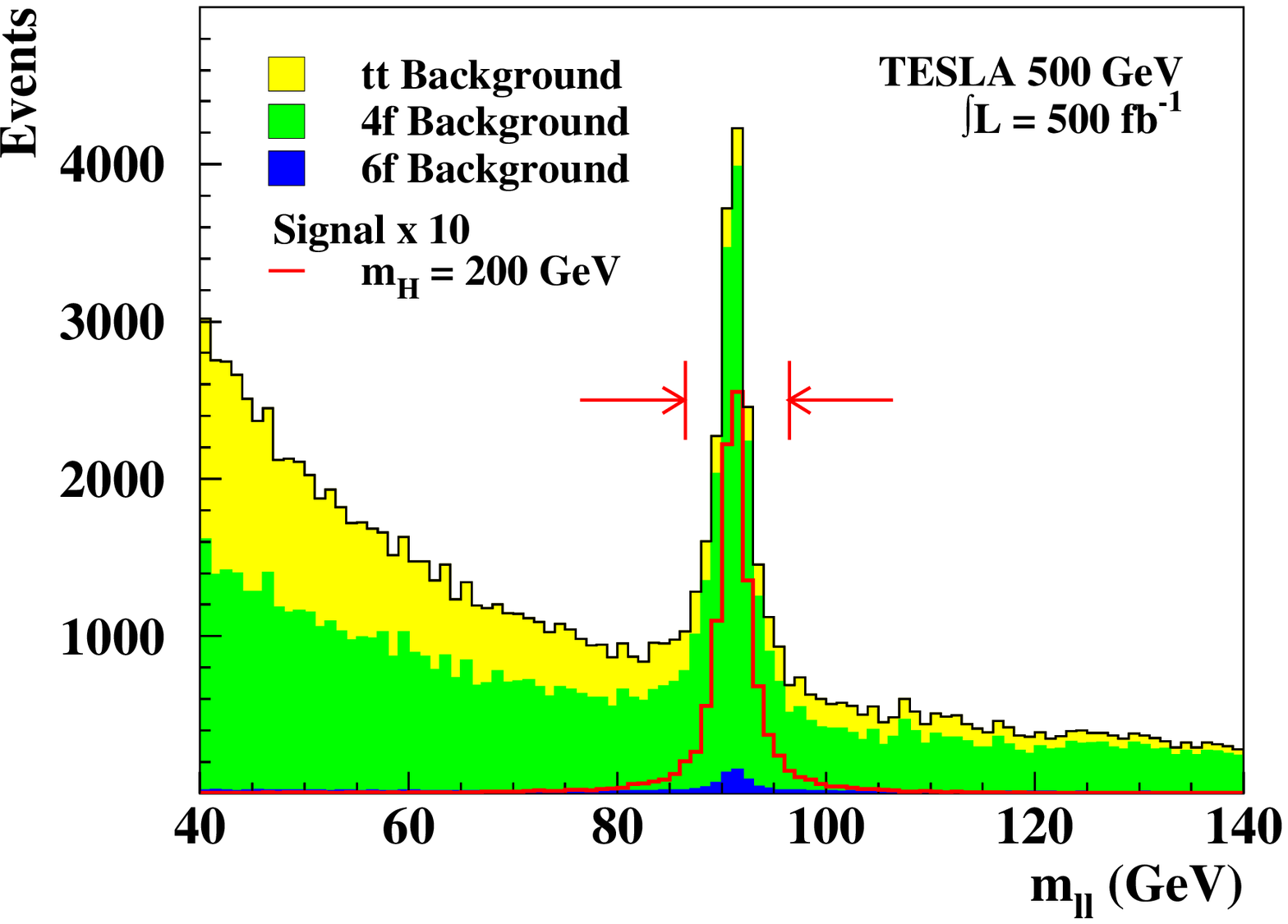,width=7.8cm}\quad\epsfig{file=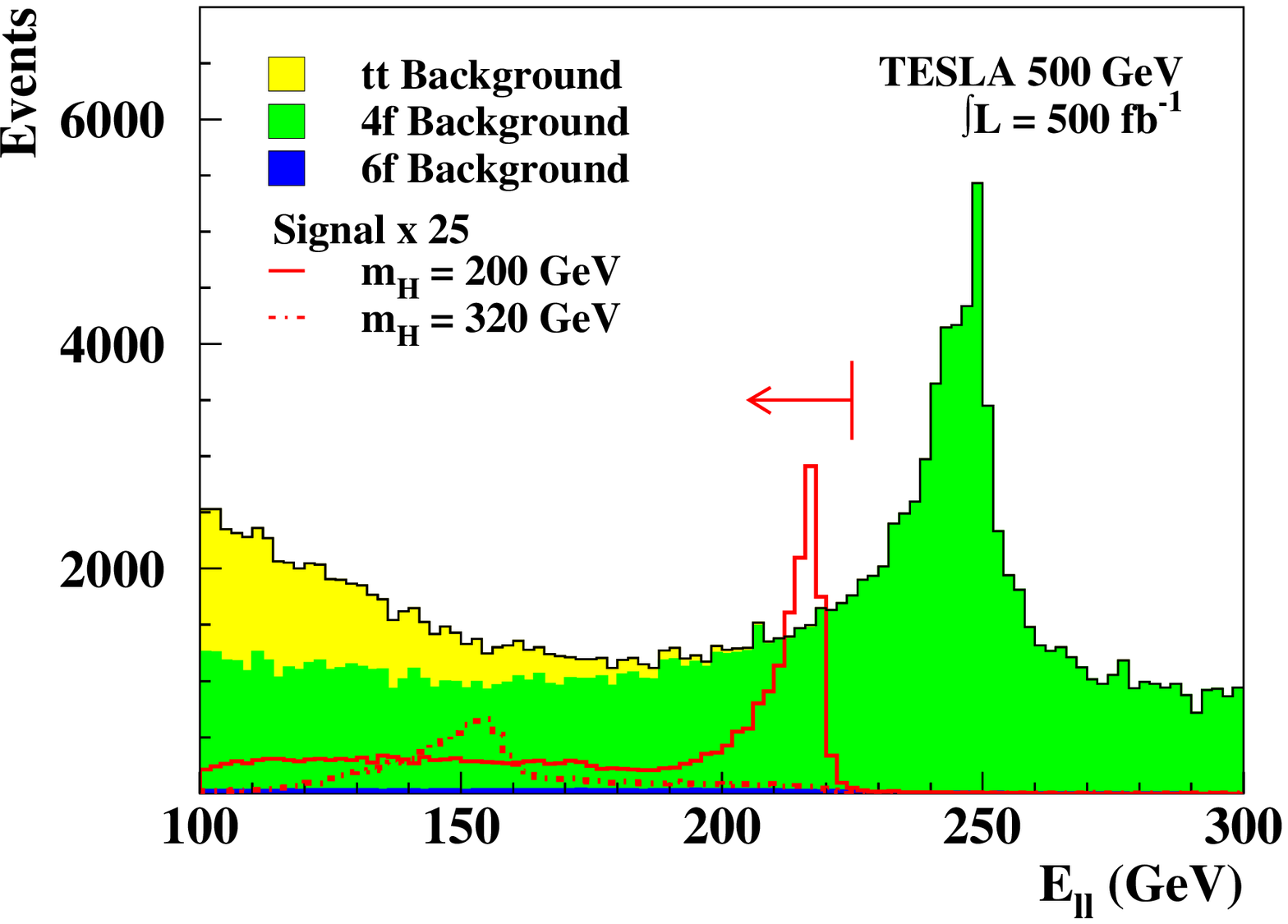,width=7.8cm}
    
    \epsfig{file=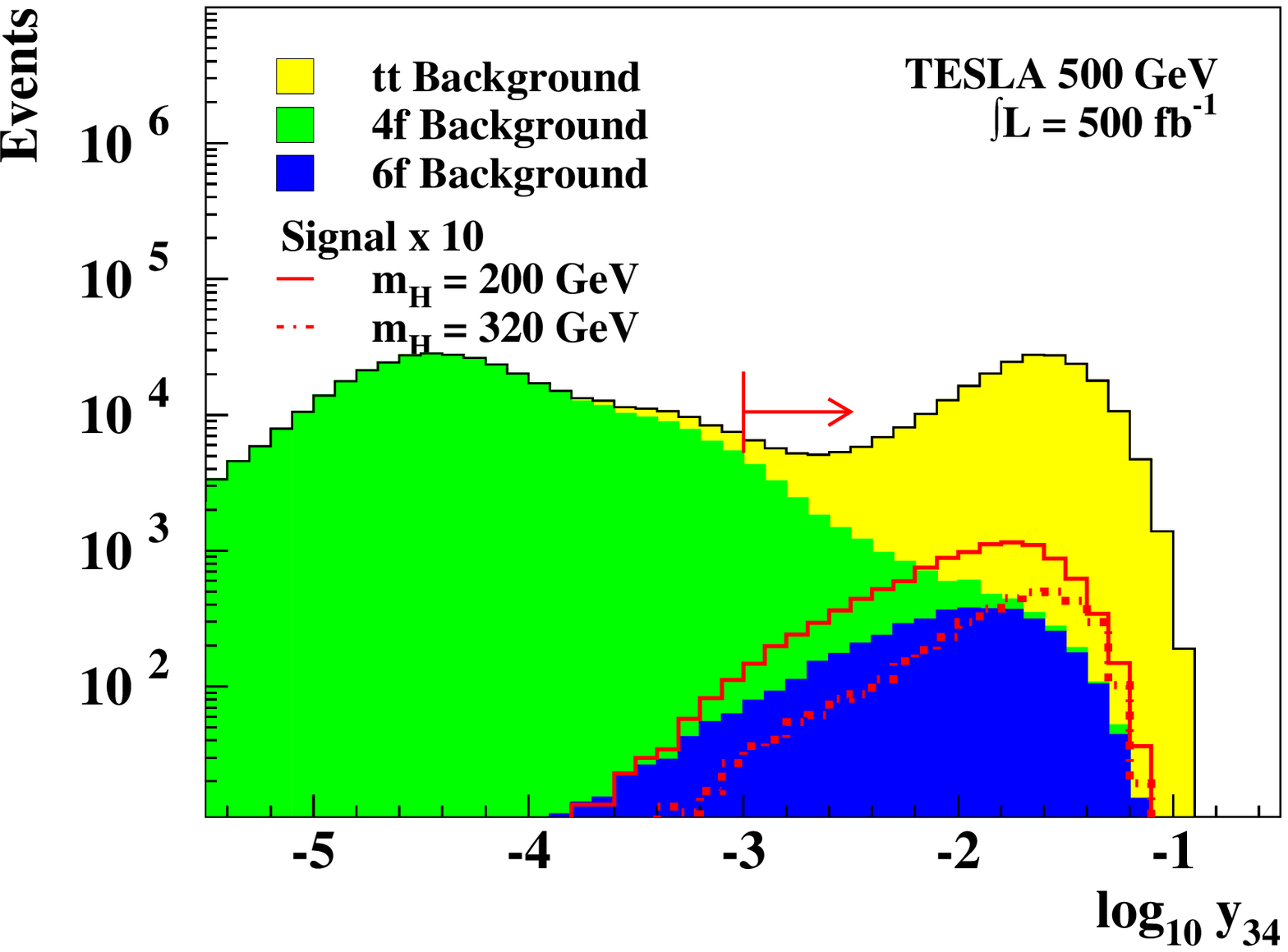,width=7.8cm}\quad\epsfig{file=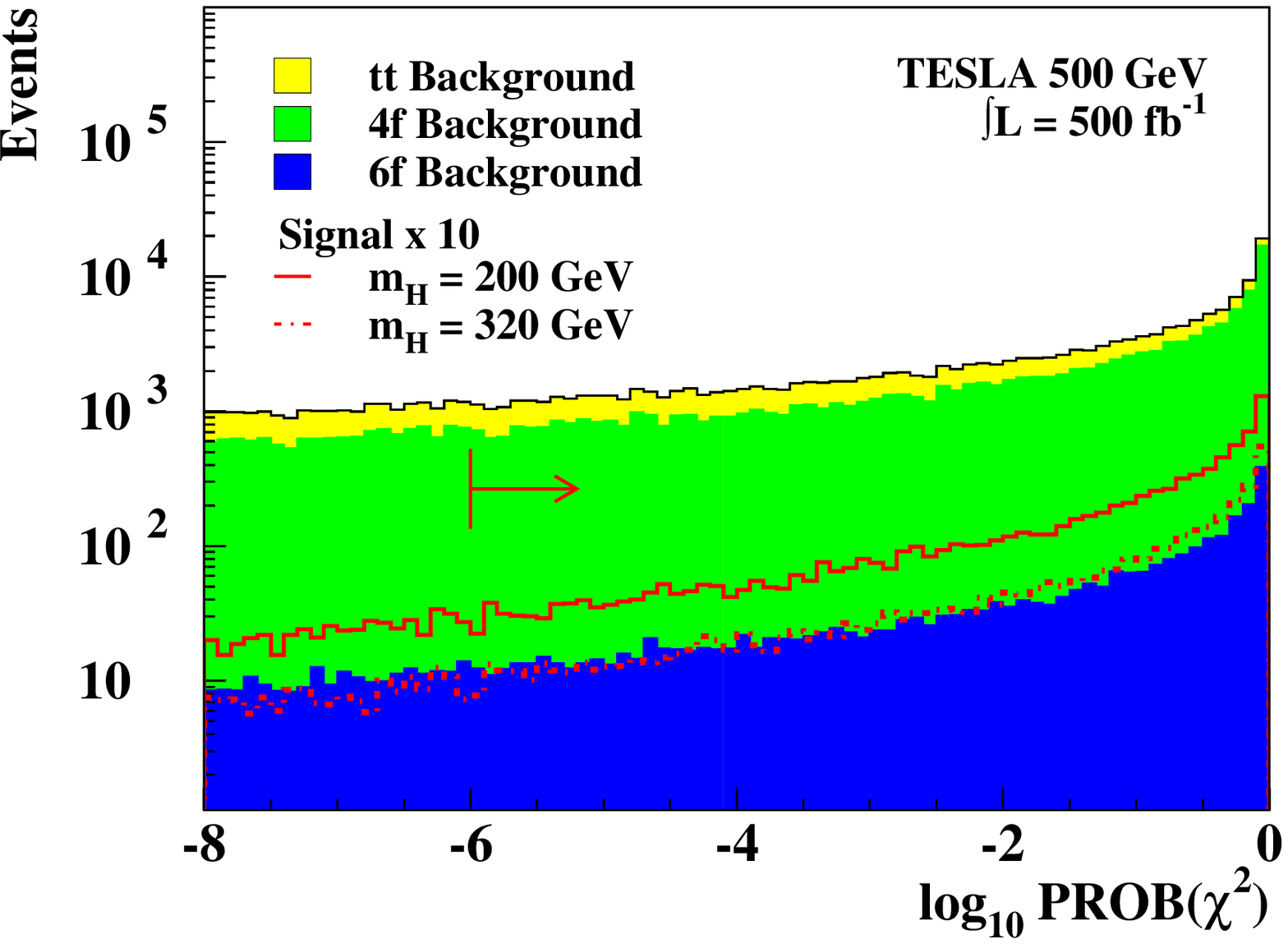,width=7.8cm}
  \caption{\label{fig:cuts}\it
    Distributions of the di-lepton mass $\mass{\lpt\lpt}$, di-lepton energy 
    $E_{\lpt\lpt}$, jet separation parameter $y_{34}$ and the
    $\chi^2$-probability of the kinematic fit $PROB(\chi^2)$. Cuts are
    indicated by arrows.}}
\end{figure}

The distributions of the most important variables used for background
suppression are displayed in Fig.~\ref{fig:cuts}.
Each event is required to pass the following cuts:
\begin{enumerate}
\item The event must contain at least two energy flow objects classified as
  electron or muon by the detector reconstruction, $N_{\lpt}\ge 2$.
\item Both leptons from the $\ZB$-decay must satisfy
  $\left|\cos\theta_{\lpt}\right|<0.99$, where $\cos\theta_{\lpt}$ is the
  lepton's polar angle.
\item All jets must satisfy $\left|\cos\theta_{\mathrm{jet}}\right|<0.95$.
\item The two lepton invariant mass must be close to the $\ZB$-mass,
  $\left|\mass{\lpt\lpt}-\mass{\ZB}\right|<5\,\GeV$. This reduces background 
  events with non-resonant lepton pairs.
\item The sum of the energy of both leptons must be significantly less than
  the beam energy, \mbox{$E_{\lpt\lpt}<225\,\GeV=0.45\,\sqrt{s}$}. In events
  from $\ZB$-pair production, each $\ZB$-boson carries half of the event
  energy, so events of this type are reduced.
\item The hadronic system must be four-jet like, $y_{34}>10^{-3}$. Here,
  $y_{34}$ is the separation parameter between three (small $y_{34}$) and four
  (large $y_{34}$) jets of the jet finder. This cut reduces most of the 
  remaining four-fermion events with non-resonant leptons still left after the 
  $\mass{\lpt\lpt}$ cut.
\end{enumerate}

In addition, a kinematic fit with four constraints is applied. The
aim is to improve the mass resolution and to group the four jets into two
pairs. The constraints are as follows: 
\begin{enumerate}
\item[1.-2.] Conservation of transverse momentum: \\
  $\sum p_x = 0$, $\sum p_y = 0$;
\item[3.] Conservation of energy and longitudinal momentum allowing for one
  initial hard photon in $\pm z$-direction: \\
  $(\sum E-\sqrt{s})^2 - (\sum p_z)^2 = 0$;
\item[4.] Four jets form two pairs of equal mass: \\
  $\mass{1,2} = \mass{3,4}$.
\end{enumerate}

This fit is performed for all three possible assignements of the four jets to
two pairs in the fourth constraint. The combination yielding the best fit
$\chi^2$ is used. Events with a probability of this $\chi^2$ below
$PROB(\chi^2)<10^{-6}$ are rejected. 
\begin{table}[htbp]
  \center{
    \begin{tabular}{|lcc||r|r|r|r||r|r|r|}
      \hline
      & & & \multicolumn{4}{|c||}{Signal $\mass{\HB}\,[\GeV]$} & \multicolumn{3}{c|}{Background} \\
      \multicolumn{3}{|c||}{Variable (range)} & 200 & 240 & 280 & 320 &
      \multicolumn{1}{c|}{6f} & \multicolumn{1}{c|}{4f} &
      \multicolumn{1}{c|}{$\tq\tqb$} \\
      \hline\hline
      \multicolumn{3}{|c||}{Events / $500\,\ifb$} & 1120 & 880 & 630 & 410 & 4340 & 392\,000 & 240\,000 \\
      \hline
      1.-3. & \multicolumn{2}{c||}{$N_{\lpt}$, $\cos{\theta_{\lpt}}$, $\cos{\theta_\mathrm{jet}}$} & 904 & 706 & 498 & 322 & 1803 & 38\,000 & 90\,000 \\
      4. & $E_{ll}$ & ($<225\,\GeV$) & 897 & 705 & 497 & 321 & 1512 & 5764 & 90\,000 \\
      5. & $\mass{ll}$ & ($\mass{\ZB}\pm 10\,\GeV$) & 770 & 600 & 425 & 275 & 369 & 1506 & 750 \\
      6. & $y_{34}$ & ($>10^{-3}$) & 745 & 581 & 413 & 268 & 343 & 5 & 357 \\
      7. & PROB($\chi^2$) & ($>10^{-6}$) & 585 & 463 & 333 & 216 & 271 & 0 & 4 \\
      \hline
      \multicolumn{3}{|c||}{Efficiency} & 52\,\% & 53\,\% & 53\,\% & 53\,\% & \multicolumn{3}{c|}{} \\
      \hline
    \end{tabular}
  \caption{\label{tab:cutflow}\it 
    Evolution of event rates through the selection. Main remaining background
    source are events from {\rm 6f} processes $\epem\to 2\lpt\,4\q$. Event
    rates in the first line include MC level cuts as described in the text.}}
\end{table}

Tab.~\ref{tab:cutflow} shows the overall performance of the event
selection. Background suppression is possible to $S/B\sim 1$ or better
depending on the Higgs mass. Events from four fermion processes and
$\tq\tqb$-pair production can be rejected almost completely. Signal efficiency
is stable as a function of $\mass{\HB}$ and lays above
$\epsilon_{signal}> 50\,\%$. 

In the following, only three objects are considered per event: A
reconstructed $\ZB$-boson (the two leptons) and two further objects assumed to
be either a pair of $\WB$- or $\ZB$-bosons (the jet pairs).

\boldmath
\section{\label{sec:det}Separation of $\hww$ and $\hzz$}
\unboldmath

Since it is {\it a priori} unknown which two of the three bosons originate
from the Higgs decay, a distribution is formed which contains all three
possible di-boson masses per event. It is expected that the correct pairing
will exhibit a mass peak while the two wrong pairings will form a flat
combinatorical background. The invariant di-boson mass spectrum is shown in
Fig.~\ref{fig:res}, where the Higgs resonance is clearly visible for all Higgs
masses.

In order to determine the resonance parameters ($\mass{}$, $\Gamma$, $N$) from
this distribution, the theoretical Breit-Wigner shape of the resonance has to
be convoluted with a properly tuned detector resolution function. The detector
resolution is estimated by using MC with zero Higgs width, 
$\Gamma_{\HB} = 0$ and the same selection as described before. 
\begin{figure}[htbp]
\center{\epsfig{file=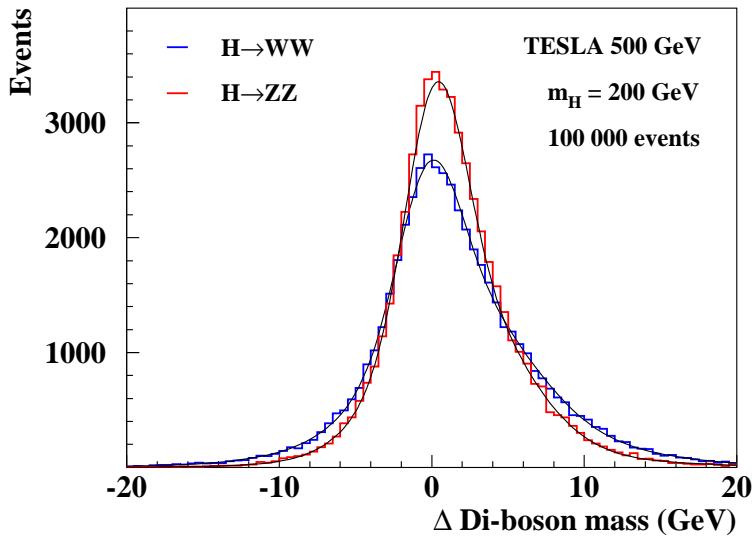,width=10cm}
\caption{\label{fig:detres}
  \it Di-boson mass resolution from MC with $\Gamma_{\HB}=0\,\GeV$. As can be
  seen, resolutions differ for $\hww$ and $\hzz$. Both distributions are 
  asymmetric due to the constraints of the kinematic fit.}}
\end{figure}

It turns out
that this detector resolution is different for $\hww$ and $\hzz$ decays, both
of which enter the di-boson mass spectrum with relative fractions 
$$\rfww = \frac{\nww}{\nww+\nzz} \qquad\mathrm{and}\qquad
\rfzz = \frac{\nzz}{\nww+\nzz}$$
respectively, with $\num{}$ being the number of events for each channel after
selection. The different resolutions arise from the fact that for
$\hzz$ the correct di-boson mass partially is an $\lpt\lpt\,\q\q$ invariant
mass, while for $\hww$ it is always $\q\q\,\q\q$. The different expected mass
spectra and the parameterizations\footnote{
Detector effects are parameterized separately for $\hww$ and $\hzz$ by
multi-gaussian functions. The choice is arbitrary and motivated by the good
agreement between MC and parameterization.} used
for convolution are shown in Fig.~\ref{fig:detres} for
$\mass{\HB}=200\,\GeV$. Resolutions for the other Higgs masses are similar.

\begin{figure}[htbp]
\center{
\epsfig{file=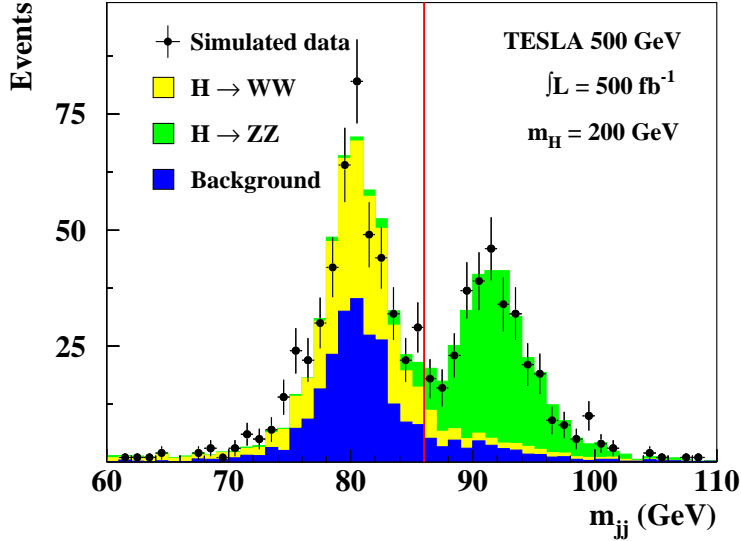,width=10cm}
\caption{\label{fig:mjj}
  \it Jet-jet mass as obtained from the kinematic
  fit. Clearly visible are two peaks at $\mass{\WB}$ and $\mass{\ZB}$
  respectively.}} 
\end{figure}

The fractions $\rfww$ and $\rfzz$ are determined from the di-jet mass as
obtained from the kinematic fit. An example for $\mass{\HB}=200\,\GeV$ is
shown in Fig.~\ref{fig:mjj}, where two peaks from $\WB$- and $\ZB$-decays
respectively are clearly visible. However, the tails are overlapping.

The di-jet mass spectrum is divided by a (in principle arbitrary) mass cut,
choosen to be $\mass{jj}=86\,\GeV$. The number of events from all event
classes ($\hww$, $\hzz$
and background), $\num{i}$ is broken up in $\num{i}^-$ events below
and $\num{i}^+$ events above this cut. The corresponding probabilities are
referred to as $\prb{i}^\pm = \num{i}^\pm /\num{i}$. Expectations for
$\prb{i}^\pm$ and $\num{bkgrd}$ are derived from MC, while the total
number $\num{tot}^\pm$ are counted in data.

From this, the relative fractions can be calculated according to:
\begin{eqnarray*}
  \num{tot}^+ & = & \num{bkgrd}^+ + \nww^+ + \nzz^+ \\
  & = & \num{bkgrd}^+ + \nww\, \pww^+ +
  \nzz\, \pzz^+ \\
  & = & \num{bkgrd}^+ + \left(\nww + \nzz\right)\, \left(\rfww\, \pww^+ +
  \rfzz\, \pzz^+\right) \\
  & = & \num{bkgrd}^+ + \left(\nww + \nzz\right)\, \left(\left(1-\rfzz \right)\,
  \pww^+ + \rfzz\, \pzz^+\right). \\
\end{eqnarray*}

With $\nww + \nzz = \num{tot} - \num{bkgrd}$ resulting in 
\begin{eqnarray*}
  \num{tot}^+ & = & \num{bkgrd}^+ + \left(\num{tot} - \num{bkgrd}\right)\, \left(\left(1-\rfzz
  \right)\, \pww^+ + \rfzz\, \pzz^+\right) \\
  \Longrightarrow\; \rfzz & = & 
  \frac{\num{tot}^+ - \num{bkgrd}^+}{\left(\num{tot}-\num{bkgrd}\right)\,
    \left(\pzz^+-\pww^+\right)} - \frac{\pww^+}{\pzz^+-\pww^+} \\
  \mathrm{and}\quad \rfww & = & 1-\rfzz = 
  \frac{\num{tot}^- - \num{bkgrd}^-}{\left(\num{tot}-\num{bkgrd}\right)\,
    \left(\pww^--\pzz^-\right)} - \frac{\pzz^-}{\pww^--\pzz^-}. \\
\end{eqnarray*}

Due to the larger background contributions below the cut, the
determination of $\rfzz$ and calculation of $\rfww = 1-\rfzz$ is more accurate
than the opposite way. Nevertheless the analogue determination of $\rfww$ by
considering events below the threshold can be used as a cross-check.

The determination of the relative fractions is model independent. In addition,
assuming there are only Higgs decays to $\WB$- or $\ZB$-pairs, the
corresponding branching rations $\brww$ and $\brzz$ can be calculated. For the
calculation, the selection efficiencies $\eww$ and $\ezz$ are
taken from MC. 

The number of events selected is a function of cross section, integrated
luminosity, selection efficiency, and branching ratios. Consequently, the
relative fraction $\rfzz$ can be expressed as
\begin{eqnarray*}
  \rfzz & = & \frac{\xsec{\HB\ZB}\,\,\lumi\,\,\ezz\,\,\brzz\,\,\br{\ZB\ZB\ZB\to
      2\lpt\,4\q}} 
     {\xsec{\HB\ZB}\,\,\lumi\,\,(\ezz\,\,\brzz\,\,\br{\ZB\ZB\ZB\to 2\lpt\,4\q} + \eww\,\,\brww\,\,\br{\ZB\WB\WB\to 2\lpt\,4\q})} \\
  & = & \frac{\ezz\,\,\brzz\,\,\br{\ZB\ZB\ZB\to
      2\lpt\,4\q}}{\ezz\,\,\brzz\,\,\br{\ZB\ZB\ZB\to 2\lpt\,4\q} +
    \eww\,\,\brww\,\,\br{\ZB\WB\WB\to 2\lpt\,4\q}}.
\end{eqnarray*}
With $\brww+\brzz=\rfww+\rfzz=1$ and $\br{\ZB\ZB\ZB\to 2\lpt\,4\q} =
3\,\br{\ZB\WB\WB\to 2\lpt\,4\q}$ this transforms to 
\begin{eqnarray*}
  \brzz & = & \frac{\eww\,\,\rfzz}{\eww\,\,\rfzz + 3\,\,\ezz\,\,\rfww}.
\end{eqnarray*}

The factor 3 arises from the threefold ambiguity in $\ZB\ZB\ZB\to 2\lpt\,4\q$.

Errors for relative fractions and branching ratios are calculated using error 
propagation. All parameters derived from MC are assumed to be known exactly, so
that $\Delta\num{tot}^\pm = \sqrt{\num{tot}^\pm}$ are the only errors taken 
into account. Values obtained are listed in Tab.~\ref{tab:rfbr}.
\begin{table}[htbp]
\center{
\begin{tabular}{|c||c|c||c|c|}
\hline
\rule[-3mm]{0mm}{8mm}$\mass{\HB}\,[\GeV]$ & $\frac{\Delta\rfww}{\rfww}$ & $\frac{\Delta\rfzz}{\rfzz}$ & $\frac{\Delta\brww}{\brww}$ & $\frac{\Delta\brzz}{\brzz}$ \\
\hline\hline
200 & $7.3\,\%$ & $5.7\,\%$ & $3.5\,\%$ & $9.9\,\%$ \\
240 & $8.9\,\%$ & $7.0\,\%$ & $5.0\,\%$ & $10.8\,\%$ \\
280 & $11.9\,\%$ & $9.3\,\%$ & $7.7\,\%$ & $16.2\,\%$ \\
320 & $15.2\,\%$ & $12.0\,\%$ & $8.6\,\%$ & $17.3\,\%$ \\
\hline
\end{tabular}
\caption{\label{tab:rfbr}\it
   Results for relative fractions and branching ratios}}
\end{table}

\section{\label{sec:mhd}Reconstruction of Higgs Resonance Parameters}

Finally, the spectrum of the reconstructed di-boson masses is built. Each
event  yields three entries for the different combinations possible. This
spectrum can be described by following parts: 
\begin{enumerate}
\item The correct combination of di-bosons form a clear peak. This peak can be
  described as a theoretical Breit-Wigner function convoluted with the
  detector resolution. The Breit-Wigner parameters $\mu$, $\Gamma$ and
  $\num{}$ are free parameters of the fit, while the parameters of the
  detector resolution are fixed from MC. 
\item The wrong combinations of di-boson masses form a flat combinatoric
  background. The shape is parameterized by a step function whose parameters
  are fixed from the same MC sample as the detector resolution\footnote{Here
    it is assumed that the shape of the combinatorical background does not
    depend on the Higgs width.} while the number of entries is determined in
  the fit by the free parameter $\num{}$ of the Higgs peak Breit-Wigner
  function.  
\item In addition, there is a flat distribution from background events left
  after the selection. This physical background is parameterized by a similar
  step function as the combinatorical background. In this case besides the
  parameters for the shape also the
  number of entries is fixed by MC expectation\footnote{In later experiments,
    the background parameters can be determined off-peak in data.}. 
\end{enumerate}

Fig.~\ref{fig:res} shows the distributions of reconstructed di-boson masses
with the parts described and the fitted function. The fit parameters
$\mu=\mass{\HB}$, $\Gamma=\Gamma_{\HB}$ and
$N=\xsec{}\times\br{}\times\epsilon$ are determined in 100 independent
MC experiments, each corresponding to $\lumi=500\,\ifb$ of integrated
luminosity. Both mean values and spread of the results lay within
statistical expectations, mean values for the measurement precision are listed
in Tab.~\ref{tab:res}. 
\begin{figure}[htpb]
\center{
\epsfig{file=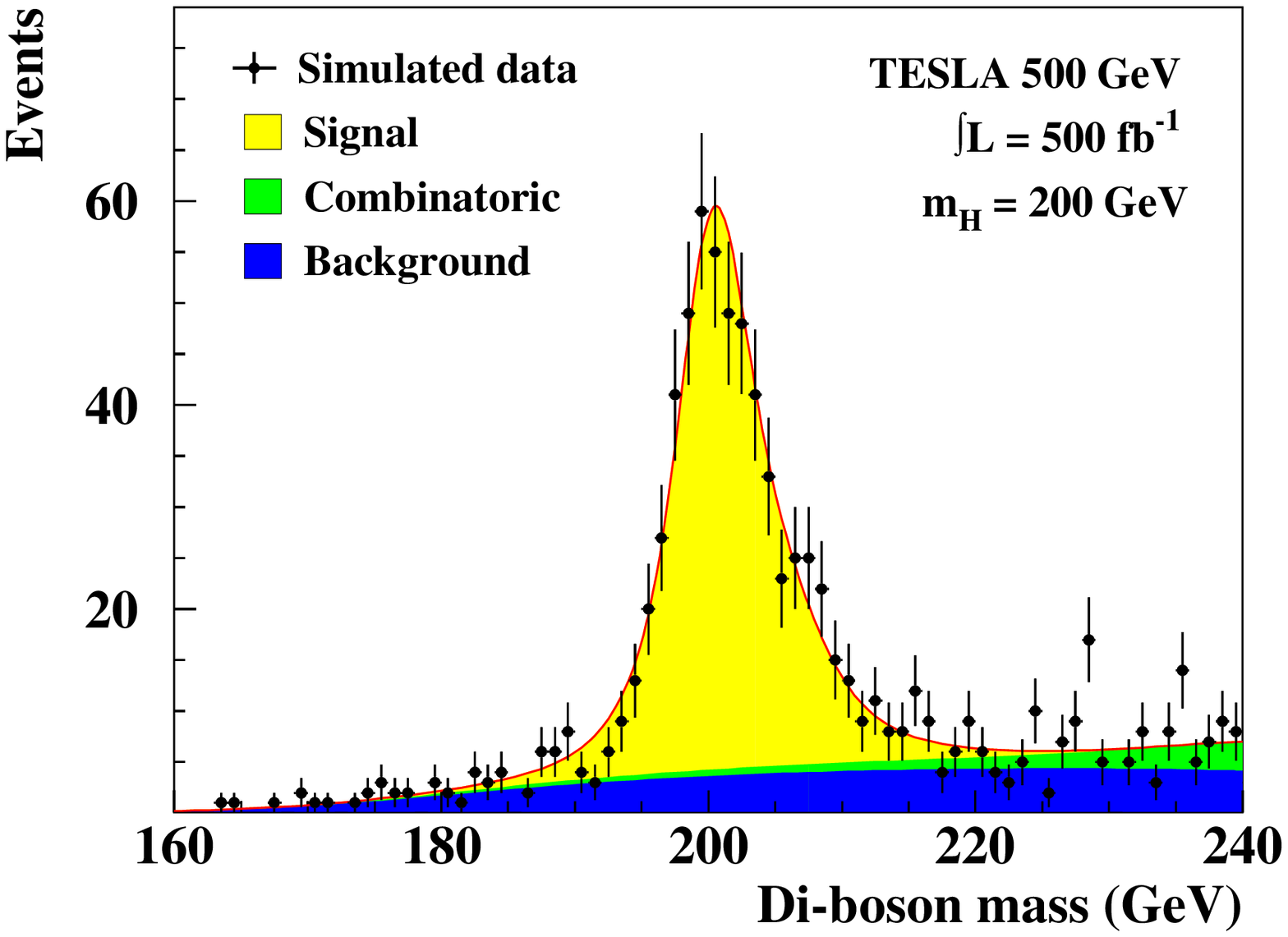,width=8cm}\epsfig{file=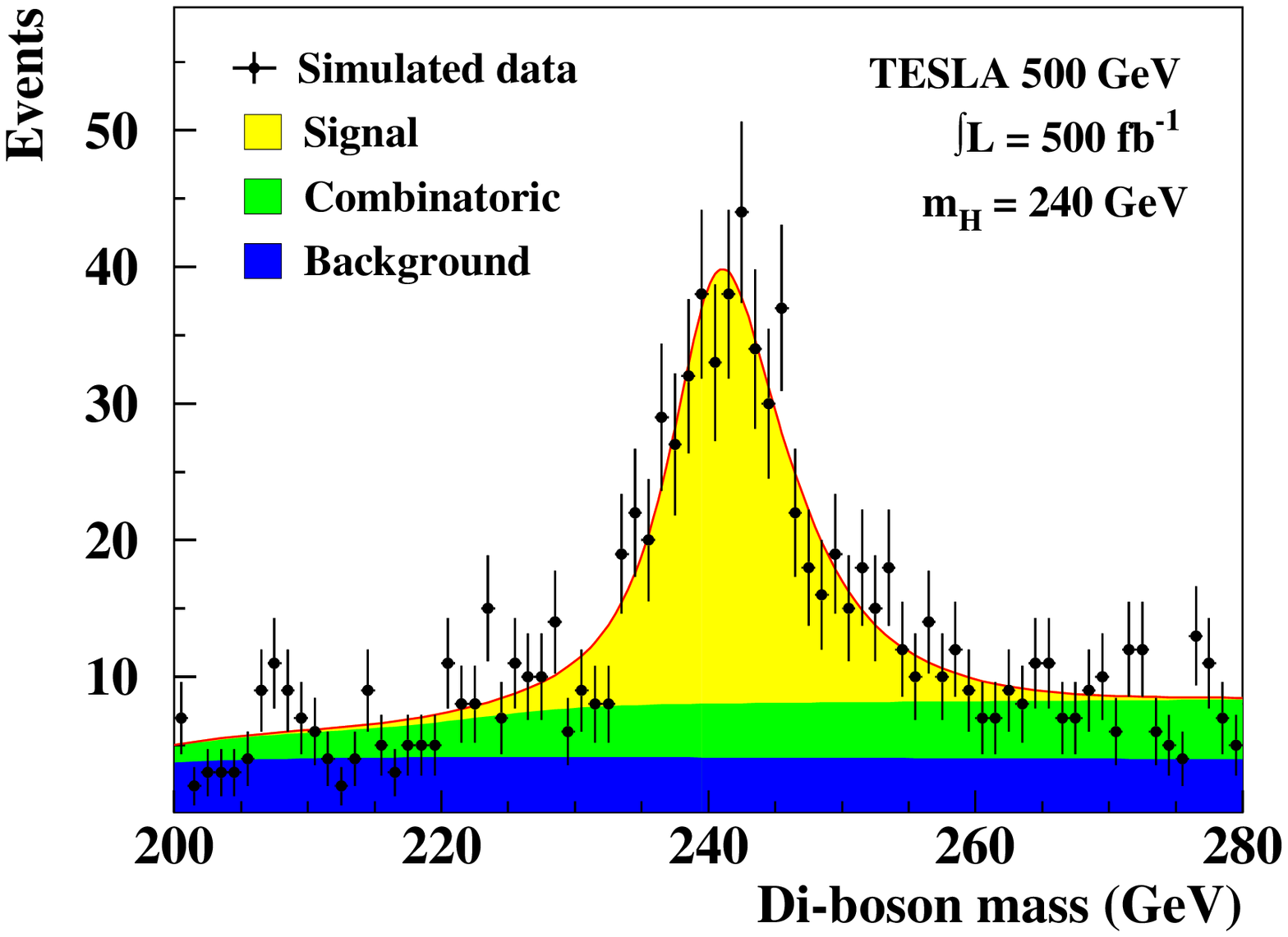,width=8cm}
\vspace{.3cm}

\epsfig{file=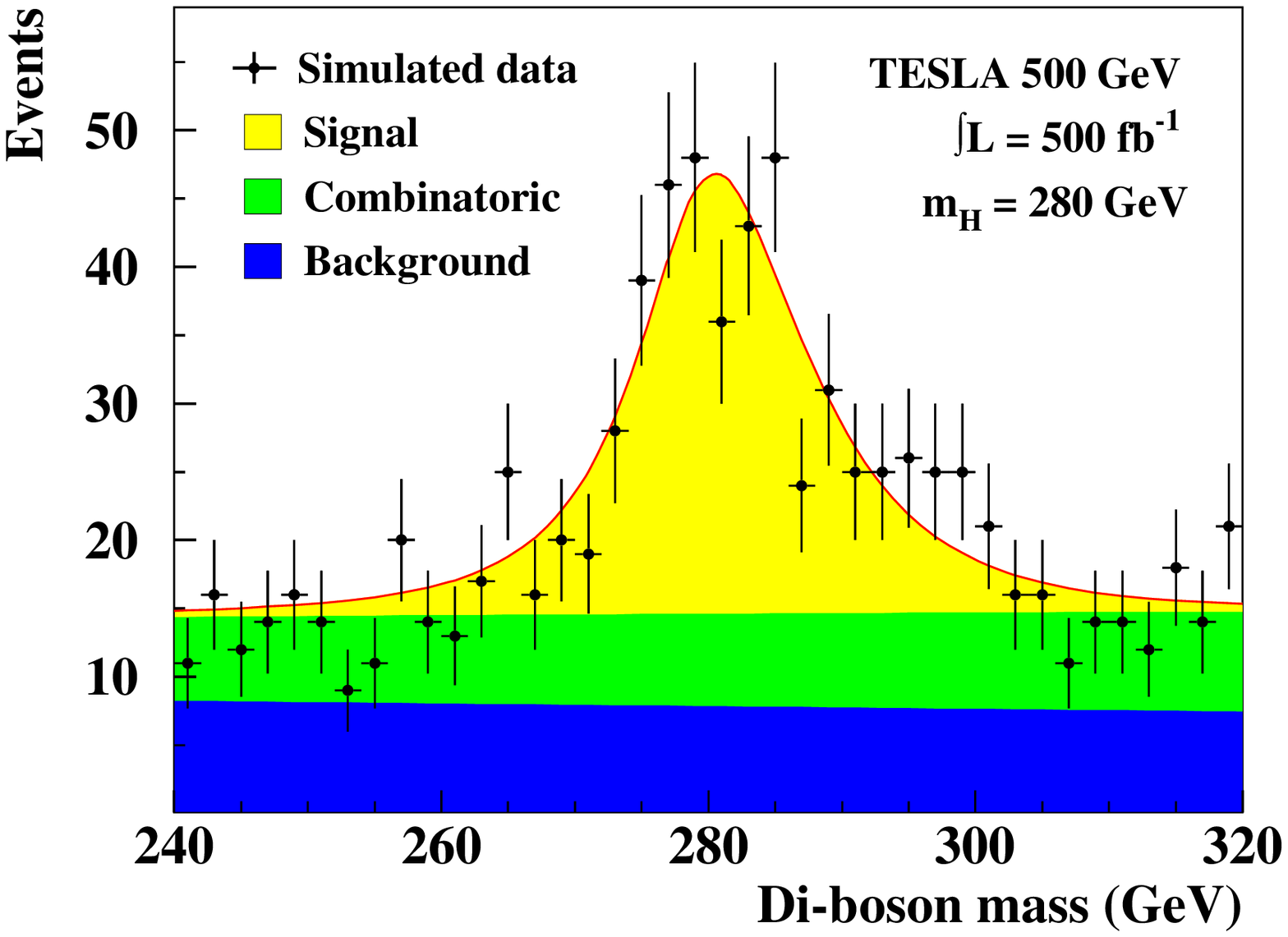,width=8cm}\epsfig{file=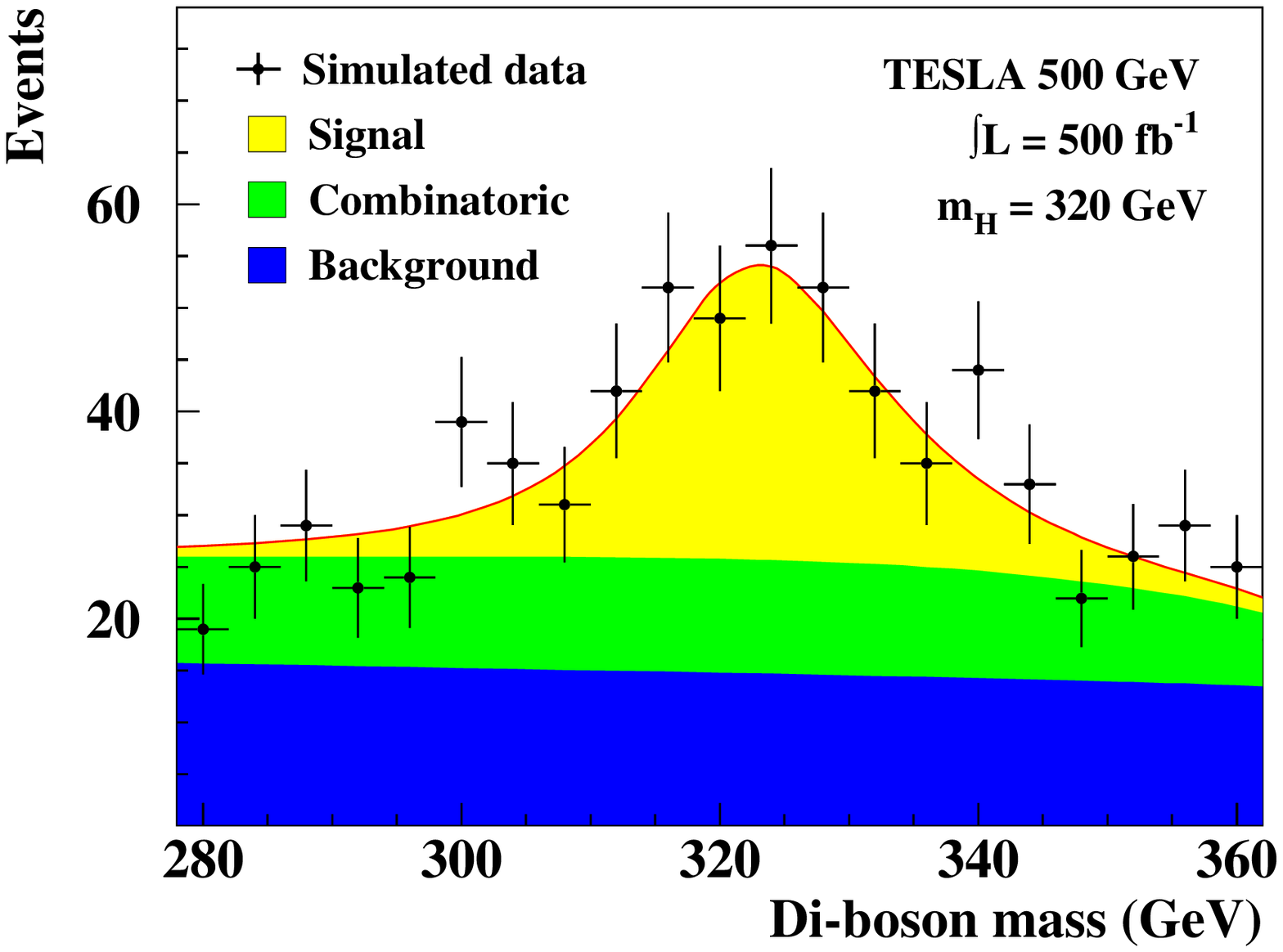,width=8cm}
\caption{\label{fig:res}\it
   Di-boson mass spectra for the four Higgs masses under study. Everywhere, 
   the higgs resonances (yellow) are clearly visible over physical (blue) 
   and combinatorical (green) background for $\lumi = 500\,\ifb$ of integrated
   luminosity at $\sqrt{s} = 500\,\GeV$.}}
\end{figure}
\begin{table}[htbp]
\center{
  \begin{tabular}{|c||c|c|c||c|c|}
    \hline
    \rule[-3mm]{0mm}{8mm}$\mass{\HB}$ & $\frac{\Delta N}{N}$ & $\frac{\Delta\mass{}}{\mass{}}$ &
    $\frac{\Delta\Gamma}{\Gamma}$ & $\frac{\Delta\brww}{\brww}$ & $\frac{\Delta\brzz}{\brzz}$ \\
    \hline\hline
    $200\,\GeV$ & $3.6\,\%$ & $0.11\,\%$ & $34.0\,\%$ & $3.5\,\%$ & $9.9\,\%$ \\
    $240\,\GeV$ & $3.8\,\%$ & $0.17\,\%$ & $26.8\,\%$ & $5.0\,\%$ & $10.8\,\%$ \\
    $280\,\GeV$ & $4.4\,\%$ & $0.24\,\%$ & $22.7\,\%$ & $7.7\,\%$ & $16.2\,\%$ \\
    $320\,\GeV$ & $6.3\,\%$ & $0.36\,\%$ & $26.4\,\%$ & $8.6\,\%$ & $17.3\,\%$ \\
    \hline
  \end{tabular}
\caption{\label{tab:res}
  \it Resolutions for event rate $N$, Higgs mass $\mass{\HB}$ and Higgs width
  $\Gamma_{\HB}$ as fitted to the di-boson mass spectrum. Also listed are
  results for the branching ratios $\brww$ and $\brzz$ as described in
  Sec. \ref{sec:det}. All numbers are mean values obtained with 100
  independent signal samples corresponding to $\lumi=500\,\ifb$ at
  $\sqrt{s}=500\,\GeV$ each.}}
\end{table}

\section{\label{sec:res}Summary and Conclusion}

We present a method for measuring Higgs mass, width and event rate in a
model independent fit from the reconstructed Higgs lineshape at TESLA. The
method is restricted to Higgs bosons with widths as large as 
few $\GeV$. Otherwise, precision is spoiled by detector mass resolution which
is in the same order. Assuming the Higgs decays to $\WB$- and $\ZB$-boson pairs
only, determination of the corresponding branching ratios $\brww$ and $\brzz$
is possible as well. 

The selection is based on reconstructing $\epem\to\HB\ZB$ events with
successive $\HB\to\WB\WB/\ZB\ZB$ decays. Selecting final states with two
charged leptons and four jets gives a good handle on background suppression
and high event rates. Since identification of $\tau$-leptons is not modelled
in the fast detector simulation used, only  electrons and muons are
considered. 

Four jets are paired to two boson candidates by a 4C kinematic fit, two
identified leptons form a 
third boson candidate. The Higgs resonance is reconstructed as di-boson
mass. Higgs event rate, mass and width are extracted from the resonance
lineshape in a model independent fit. The results of 100 independent MC
experiments each corresponding to $\lumi=500\,\ifb$ of integrated luminosity
lay within statistical expectation. Mean values of the precision achieved are
listed in Tab.~\ref{tab:res}. 

Main motivation for this study was to explore a direct method for measuring
the total Higgs width $\Gamma_{\HB}$. Before, this has only be studied for LHC
experiments \cite{sopczak}. Results obtained in this study are comparable in
precision for mass and width of the resonance. No LHC numbers are available
for event rates, and branching ratio determination is impossible from the
lineshape reconstruction.

Fig.~\ref{fig:exph} (left) compares the results obtained in this study on
Higgs width and mass with previous TESLA studies on lower Higgs masses. For the
Higgs width, the indirect determination via the Higgs coupling to $\WB$-bosons,
measured in the cross section of $\WB\WB$-fusion \cite{nm_fus}, is the most
precise method studied so far. However, up to now only $\HB\to\bq\bqb$ decays
have been investigated, so precision breaks down as does the branching ratio
$\br{\HB\to\bq\bqb}$ at $\mass{\HB}\gsim 150\,\GeV$. On the other hand,
precision for direct width determinations in the Higgs lineshape are limited
by the narrow Higgs width below $\mass{\HB}\lsim 200\,\GeV$. The gap could be
closed by indirect determinations with the analysis of $\HB\to\WB\WB$
decays. The hope is, that combination of direct and indirect measurements
significantly improves the precision on the Higgs width for high Higgs masses.

\begin{figure}[htbp]
\center{
\epsfig{file=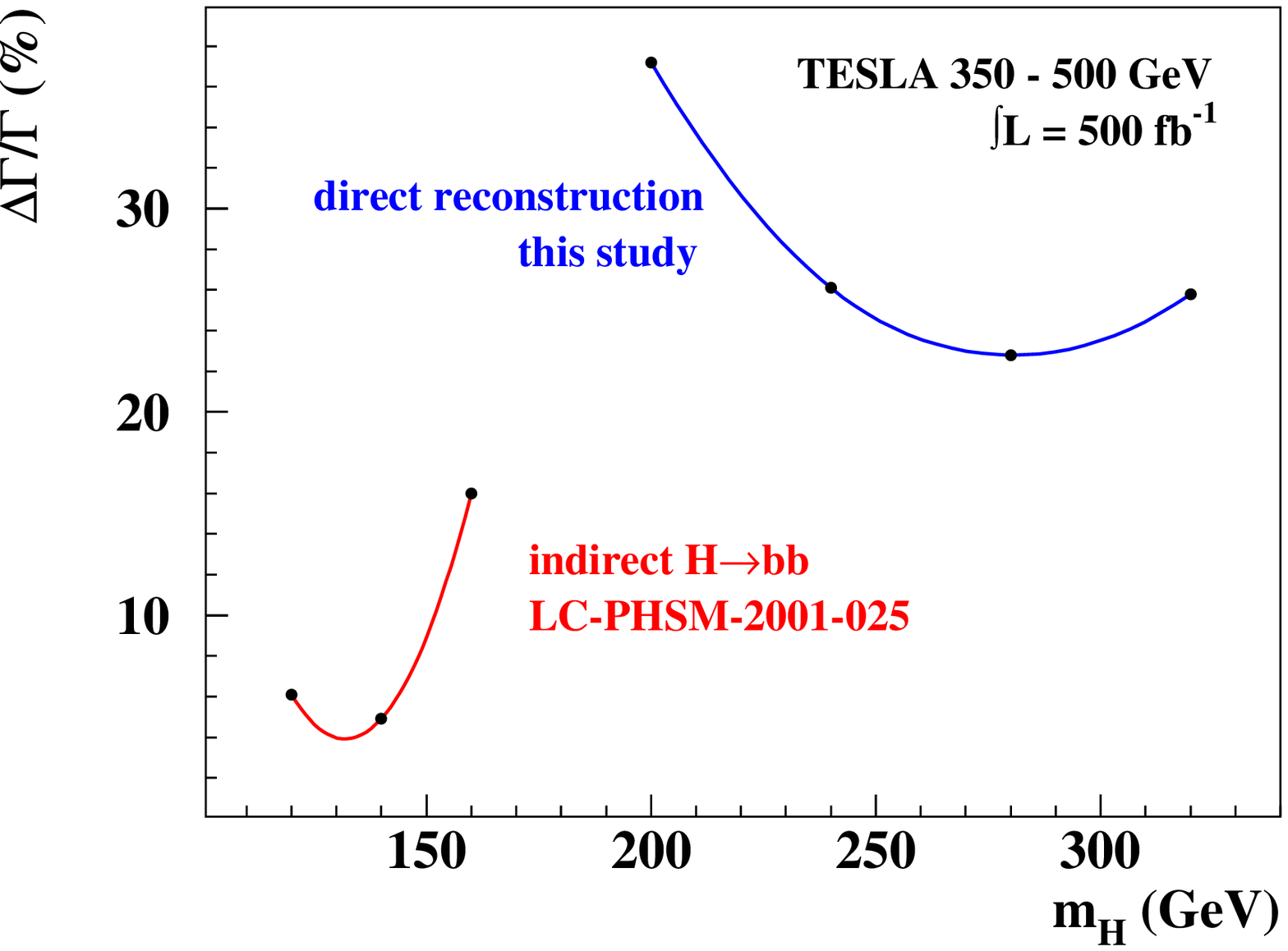,width=8cm}\epsfig{file=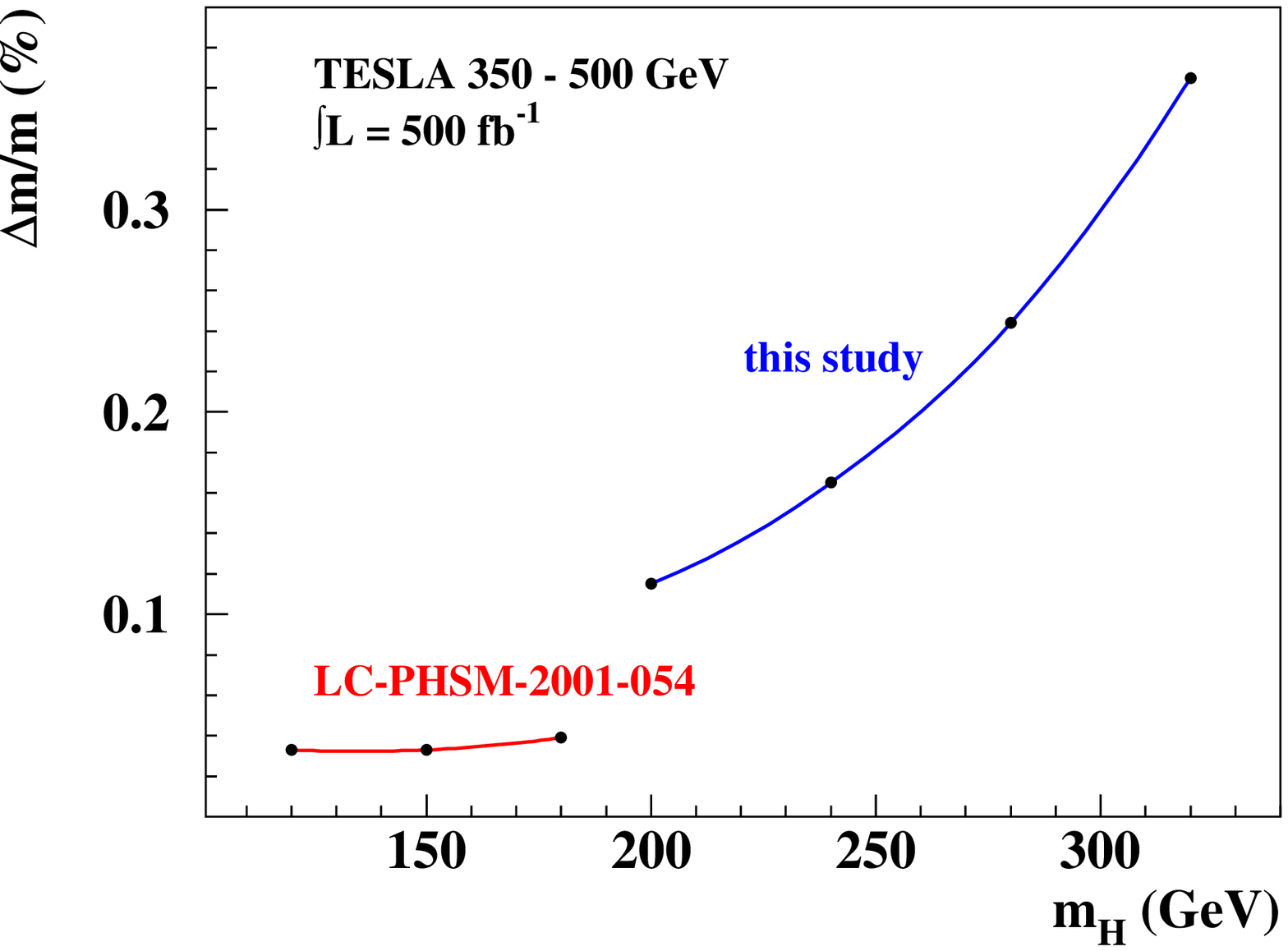,width=8cm}
\caption{\label{fig:exph}
  \it Precisions of Higgs width (left) and mass (right) measurements at
  TESLA. Results from former studies are shown in red, those of this simulation
  in blue.}}
\end{figure}

Precisions of the direct method alone may be optimized as well. There are
many ways to enhance signal event rates by considering more common final
states. For instance, one could include $\ZB\to\taup\taum$ or study
$\HB\ZB\to\q\q\q\q\q\q$ and $\HB\ZB\to\q\q\q\q\lpt\nu$ final states. But, the
effects of larger background contributions and final state neutrinos need
dedicated studies. 

It is pointed out that this study is optimized for width
determination. Results on mass, event rate and branching ratios are welcome
spin-offs, but there might be other processes and methods still to be studied
which are more appropriate. As an example, the precision on Higgs mass
measurements of previous TESLA studies \cite{rasp_mass} is compared to results
of this study in  Fig.~\ref{fig:exph} (right). As can be seen, precision is
about three times better in the dedicated study. The main reason, is 
selection of more common signal final states and thus higher statistics.

\section*{Acknowledgements}

I would like to thank all members of the FLC group at DESY for their support
and fruitful discussions on this work. I am especially grateful to Klaus Desch
and Rolf-Dieter Heuer for their support and patience -both professional and
private- during the last year.


\begin{thebibliography}{}
\bibitem{tdr}
  TESLA Technical Design Report, DESY-2001-011.

\bibitem{Abe:2001wn}
T.~Abe {\it et al.}  [American Linear Collider Working Group Collaboration],
{\it Linear collider physics resource book for Snowmass 2001},
Proc. of the APS/DPF/DPB Summer Study on the Future of Particle Physics (Snowmass 2001), ed. N.~Graf,
SLAC-R-570.

\bibitem{Abe:2001gc}
K.~Abe {\it et al.}  [ACFA Linear Collider Working Group Collaboration],
{\it Particle physics experiments at JLC},
arXiv:hep-ph/0109166.

\bibitem{hzha}
  P.~Janot, 
  {\it Physics at LEP2}, 
  CERN 96-01, Vol.2, 309.

\bibitem{hdecay}
  A.~Djouadi, J.~Kalinowski and M.~Spira, 
  {\it HDECAY: a Program for Higgs Boson Decays in the Standard Model and its
    Supersymmetric Extensions}, 
  \CPC{\bf 108}{1998}{56}.

\bibitem{ssr_gam}
  G.~Jikia, S.~S\"oldner-Rembold, 
  {\it Light Higgs Production at a Photon Collider},
  Proceedings of the 7th International Workshop on High Energy Photon
  Colliders, Hamburg 2000, 133.

\bibitem{nm_fus}
  K.~Desch, N.~Meyer, 
  {\it Study of Higgs Boson Production through WW-fusion at TESLA}, 
  {LC-PHSM-2001-025}.

\bibitem{whiz}
  W.~Kilian, 
  {\it WHiZard 1.22 - Manual}, 
  LC-TOOL-2001-039.

\bibitem{pythia}  
  T.~Sjostrand, L.~Lonnblad and S.~Mrenna,
  {\it PYTHIA 6.2: Physics and manual},
  arXiv:hep-ph/0108264.

\bibitem{circe} 
  T.~Ohl,
  {\it CIRCE version 1.0: Beam spectra for simulating linear collider physics},
  \CPC{\bf 101}{1997}{269}.

\bibitem{simdet}  
  M.~Pohl and H.~J.~Schreiber,
  {\it SIMDET - Version 4: A parametric Monte Carlo for a TESLA detector},
  arXiv:hep-ex/0206009.

\bibitem{durham} Y.~Dokshitzer, J.Phys.~{\bf G17}, (1991), 1537; \\ N.~Brown, W.~J.~Stirling, Phys. Lett. {\bf B252} (1990) 657; \\ S.~Bethke~et~al., Nucl. Phys. {\bf B370} (1992) 310;\\ S.~Catani~et~al., Phhys.~Lett.~{\bf B269} (1991),432; \\ N.~Brown, W.~J.~Stirling, Z. Phys. {\bf C53} (1992) 629.

\bibitem{sopczak}
  V.~Drollinger and A.~Sopczak,
  Eur.\ Phys.\ J.\ directC {\bf 3} (2001) N1.

\bibitem{rasp_mass} 
  P.~Garcia-Abia, W.~Lohmann and A.~Raspereza,
  {\it Measurement of the Higgs Boson Mass and Cross Section with a Linear
    $e^+e^-$ Collider},
  LC-PHSM-2001-054.

\end{thebibliography}
\end{document}